\documentclass[aip,jcp,reprint]{revtex4-2}
\usepackage{amsmath}
\usepackage{bm}
\usepackage{dsfont}
\usepackage[left = 1.35cm, right = 1.35cm, top = 1.96cm, bottom = 1.96cm]{geometry}

\usepackage{graphicx}

\usepackage{color}

\definecolor{OliveGreen}{rgb}{0.1, 0.4, 0.1}

\usepackage{setspace}
\emergencystretch=\maxdimen
\allowdisplaybreaks

\usepackage[colorlinks=true,citecolor=OliveGreen,linkcolor=OliveGreen,urlcolor=OliveGreen]{hyperref}

\definecolor{awesome}{rgb}{1.0, 0.13, 0.32}

\usepackage{mathrsfs}

\newcommand{\G}{\bm{\mathcal{G}}}
\renewcommand{\P}{\bm{\mathcal{P}}} 
\newcommand{\boldnabla}{\boldsymbol{\nabla}}
\newcommand{\boldepsilon}{\boldsymbol{\epsilon}}
\newcommand{\R}{\bm{r}}
\newcommand{\etaS}{\eta_\mathrm{S}}
\newcommand{\etaD}{\eta_\mathrm{D}}
\newcommand{\etaO}{\eta_\mathrm{O}}
\newcommand{\etaB}{\eta_\mathrm{B}}

\begin{document}

\title{Analytical response functions for a compressible thin fluid layer with odd viscosity}

\thanks{Article contributed to the topical issue of the \textit{New Journal of Physics} on ``Broken symmetries and odd transport in statistical physics'' edited by Erik Kalz, Ralf Metzler, and Abhinav Sharma.}

\author{Abdallah Daddi-Moussa-Ider}%
\email{admi2@open.ac.uk}
\thanks{Corresponding author}
\affiliation{School of Mathematics and Statistics, The Open University, Walton Hall, Milton Keynes MK7 6AA, United Kingdom}

\author{Yuto Hosaka}%
\email{yuto.hosaka@ds.mpg.de}

\affiliation{Max Planck Institute for Dynamics and Self-Organization (MPI--DS), Am Fa\ss berg 17, 37077 G\"{o}ttingen, Germany}

\affiliation{Department of Mathematics, Kyoto University, Kyoto 606-8502, Japan}

\date{\today}%

\author{Shigeyuki Komura}
\email{komura@wiucas.ac.cn}
\affiliation{Zhejiang Key Laboratory of Soft Matter Biomedical Materials, Wenzhou Institute, University of Chinese Academy of Sciences, Wenzhou, Zhejiang 325000, China}

\begin{abstract}
 Fluids composed of chiral active components can exhibit odd viscosity, a property that breaks time-reversal and parity symmetries. We investigate the hydrodynamic response to monopole and dipole singularities in a compressible thin fluid layer with odd viscosity, supported by a conventional lubrication layer. Using the two-dimensional Green’s function in Fourier space, we derive analytical solutions for the flow and pressure fields. These solutions provide a detailed description of the hydrodynamic interactions governing the motion of colloidal particles and microswimmers in confined chiral fluids, offering insight into the role of odd viscosity in modifying particle dynamics and collective behavior. The derived results are directly applicable to modeling transport, control, and self-organization phenomena in active and chiral microfluidic systems.
\end{abstract}

\maketitle

\section{Introduction}

Chiral active matter, ranging from driven spinner collectives~\cite{soni2019odd, yang2021topologically, zhao2021, lopez2022chirality, mecke2023simultaneous, shen2023collective,
mecke2024chiral, katuri2024control, zhou2025experimental} to active bacterial suspensions~\cite{beppu2021edge, li2024robust}, naturally breaks time-reversal and parity symmetries~\cite{shankar2022topological}. 
In these non-equilibrium states, systems can exhibit a novel transport coefficient known as odd or Hall viscosity~\cite{avron1998}.
This is a non-dissipative, antisymmetric contribution to the fluid viscosity tensor that drives flows transverse to the velocity gradient~\cite{banerjee2017, liebchen2022chiral, hosaka2022nonequilibrium, fruchart2023odd, mecke2024emergent, sone2026hermitian}. 
In quasi-two-dimensional (2D) setups, such behavior clearly emerges under compressibility conditions and leads to a variety of remarkable transport phenomena, such as topological waves at fluid boundaries~\cite{souslov2019topological}, and nonreciprocal and transverse hydrodynamic responses~\cite{hosaka2021nonreciprocal, hosaka2023pair, lier2023lift, mecke2023simultaneous}.
Consequently, the linear relation between hydrodynamic forces and the resulting velocities is no longer constrained by classical Onsager reciprocity~\cite{onsager1931reciprocal}, which is expressed as the Lorentz reciprocal theorem in low-Reynolds-number fluid dynamics~\cite{fruchart2023odd, hosaka2023lorentz}. It is important to note, however, that while standard reciprocity is broken in the presence of odd viscosity, the transport remains subject to the generalized constraints of the Onsager-Casimir relations~\cite{onsager1931reciprocal2, casimir1945onsager}, which have been reported commonly in systems with odd viscosity~\cite{fruchart2023odd}.

Compressible fluids with odd viscosity are often modeled as thin fluid layers with momentum leakage to a surrounding bulk or substrate~\cite{hosaka2021nonreciprocal, lier2023lift}. 
Under the incompressibility condition, on the other hand, odd viscosity does not manifest itself in 2D fluids with no-slip boundary conditions~\cite{ganeshan2017, hosaka2021hydrodynamic}.
However, it produces pronounced effects in three-dimensional (3D) bulk fluids.
Examples include transverse or lift forces~\cite{khain2022, yuan2023stokesian,
aggarwal2023thermocapillary,
hosaka2023lorentz, everts2024dissipative, khain2024trading}, chirotactic response of microswimmers~\cite{hosaka2024chirotactic}, translation-rotation coupling of moving objects~\cite{lier2024odd, w6pg-4471}, and pattern formation in turbulence~\cite{de2024pattern}.
Importantly, the notion of transverse stresses induced by odd viscosity has also been applied beyond its original setting to include other odd transports, such as odd-elastic materials~\cite{lin2024emergence, veenstra2025adaptive} and non-conservative responses~\cite{caporusso2024phase, wang2024condensate, gu2025emergence, caprini2025bubble, luigi2025self}.

In the study of hydrodynamics with odd viscosity, the central theoretical foundation is the determination of the fundamental solution of the Stokes equations with odd viscosity, i.e., the Green’s function. 
This solution forms the basis for constructing resistance tensors, multipole expansions, and boundary integral methods~\cite{pozrikidis1992}. 
Early contributions in this research area focused on the linear hydrodynamic response of a 2D compressible odd-viscous layer~\cite{hosaka2021nonreciprocal}. 
These studies derived the Green’s function semi-analytically, revealing how odd viscosity introduces an antisymmetric component to the hydrodynamic propagator and generates perpendicular flow components.

Subsequent works have explored the dynamical implications of the Green's function in more complex settings~\cite{duclut2024probe}, including analyses of the lift force on a translating disk in odd viscoelastic media. More recently, some of the present authors employed a 2D Fourier transform approach to analytically determine the velocity and pressure fields around a disk-shaped inclusion in a supported fluid layer, enabling estimation of the resistance coefficients for a disk in motion~\cite{daddi2025analytical, daddi2025hydrodynamic}. Meanwhile, the hydrodynamics of model self-propelled microswimmers in chiral fluids has been investigated~\cite{hosaka2023hydrodynamics}, showing that a pair of swimmers exhibits a rich spectrum of two-body dynamics, governed by their initial relative orientations and the specific propulsion mechanisms of each swimmer.

Despite these recent advances, the analytical linear response functions of a 2D compressible fluid layer with odd viscosity, together with conventional shear and dilational viscosities, remain unresolved. This has prevented a thorough  understanding of the more complex transport phenomena and multi-body dynamics, which have only been studied using an approximate Green's function in the limit of small odd viscosity~\cite{hosaka2023hydrodynamics}. The primary challenge arises from the nontrivial coupling between the degrees of freedom introduced by odd viscosity and the compressibility of the fluid layer.

In this work, we derive the exact Green’s functions using a 2D Fourier transform formulation. Building on these results, we determine the hydrodynamic fields generated by monopole and dipole singularities and analyze the transverse flow response induced by odd viscosity. We find that the interplay between the 2D fluid layer and the underlying 3D hydrodynamics gives rise to rich flow structures, including vortical patterns. The hydrodynamic screening lengths that characterize these structures are determined solely by the conventional (even) viscosities, while the emergent transverse response is governed exclusively by the odd-viscosity coefficient.

\begin{figure}
    \centering
    \includegraphics[width=\linewidth]{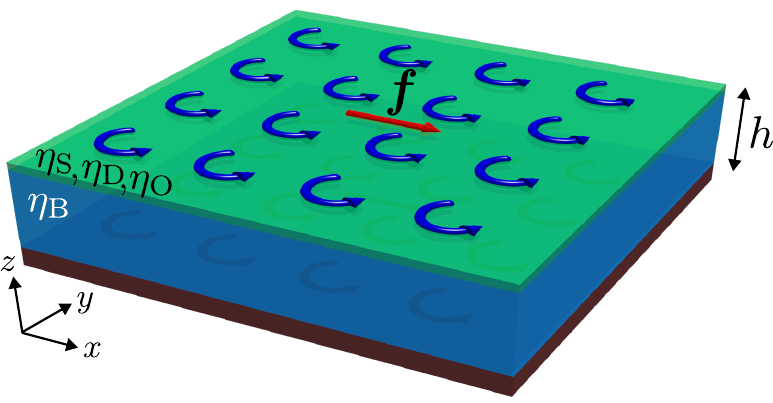}
    \caption{
    Schematic illustration of a 2D chiral active fluid layer, which breaks time-reversal and parity symmetries (e.g., due to internally actuated spinning constituents, as depicted by the blue arrows, or the intrinsic handedness of the fluid medium). The layer is modeled as an infinitely thin, compressible 2D fluid with 2D shear viscosity~$\etaS$, dilatational viscosity~$\etaD$, and odd viscosity~$\etaO$. 
    The fluid layer rests on a 3D bulk fluid of thickness $h$ (not to scale) and 3D shear viscosity~$\etaB$, which is bounded from below by a flat, impermeable rigid substrate. The red arrow indicates a point force~$\bm{f}$ applied to the 2D fluid layer.}
    \label{fig:system}
\end{figure}

\section{Governing equations}

We consider a 2D compressible fluid layer supported by an incompressible 3D bulk fluid, which is confined by a rigid wall and has 3D shear viscosity~$\etaB$. 
We denote the depth of the underlying 3D fluid by~$h$; see Fig.~\ref{fig:system} for a schematic illustration of the system. 
We define the position vector on the 2D fluid layer as $\bm{r} = (x, y)$.
The momentum balance for the 2D fluid layer is governed by
\begin{equation}
    \boldnabla_\parallel \cdot \boldsymbol{\sigma} + \bm{f}_{\rm B} + \bm{f} = \mathbf{0} \, , 
    \label{eq:momentum_2D}
\end{equation}
where $\boldnabla_\parallel$ denotes the 2D gradient operator, $\boldsymbol{\sigma}$~is the hydrodynamic stress tensor, $\bm{f}_{\rm B}$~is the bulk force applied by the underlying bulk fluid, and $\bm{f}$~represents any additional force density exerted on the 2D fluid.

The model setup introduced above — a compressible fluid layer in contact with an underlying 3D fluid — provides a generic framework for exploring the properties of a 2D odd viscosity. 
In a purely 2D incompressible fluid, the odd viscosity term in the momentum balance is unobservable because it acts as a gradient of a scalar, which can be absorbed into the pressure field~\cite{ganeshan2017}. 
To reveal its elusive effects, one must therefore introduce mechanisms that circumvent this constraint, for example, fluid compressibility due to hydrodynamic coupling with a 3D bulk fluid, as adopted here. 
This 2D-3D coupling is ubiquitous in soft matter, such as soluble surfactant monolayers in contact with 3D fluids~\cite{barentin1999, elfring2016surface}, and serves as a basis for modeling 2D fluids with odd viscosity~\cite{hosaka2021nonreciprocal, lier2023lift}.

The bulk force $\bm{f}_{\rm B}$ exerted by the underlying bulk fluid on the 2D fluid layer is obtained by solving the corresponding hydrodynamic equations~\cite{barentin1999}. The bulk fluid is governed by the 3D incompressible Stokes equations
\begin{equation}
    \eta_{\rm B} \boldnabla^2 \bm{u} - \boldnabla p = \mathbf{0} \, , 
    \quad    
    \boldnabla \cdot \bm{u} = 0 \, , 
\end{equation}
wherein $\bm{u}$ and $p$ denote the 3D velocity and pressure fields, respectively, and $\etaB$ the shear viscosity of the underlying 3D fluid.
The 3D flow satisfies no-slip boundary conditions at the impermeable rigid wall at $z = 0$, i.e., $\bm{u}(\bm{r}, 0) = \bm{0}$; see Fig.~\ref{fig:system}.
Additionally, continuity of velocity at $z = h$ requires $\bm{u}(\bm{r}, h) = \bm{v}(\bm{r})$, where $\bm{v}(\bm{r})$ is the velocity of the 2D fluid layer.
By using the lubrication approximation, which is valid when the layer thickness~$h$ is much smaller than any lateral dimension, the bulk velocity is expressed as~\cite{barentin1999}
\begin{equation}
    \bm{u}(\bm{r}, z) = \frac{1}{2\etaB}\, z (z-h) \, \boldnabla_\parallel p + \frac{z}{h} \, \bm{v} \, .
\end{equation}
The force applied by the 3D bulk on the 2D fluid layer is obtained by projecting the hydrodynamic traction onto the horizontal $(x, y)$ plane as~\cite{barentin1999} 
\begin{equation}
    \bm{f}_{\rm B} = -\frac{h}{2} \, \boldnabla_\parallel p - \frac{\etaB}{h} \, \bm{v} \, .
    \label{eq:bulk_force}
\end{equation}

The viscous stress tensor in Eq.~\eqref{eq:momentum_2D} for the 2D compressible fluid with odd viscosity is given by~\cite{hosaka2021nonreciprocal}
\begin{equation}
\sigma_{ij} = \sigma_{ij}^\mathrm{E}
+ \sigma_{ij}^\mathrm{O} \, ,
\label{eq:stress_tensor}
\end{equation}
where
\begin{subequations}
    \begin{align}
    \sigma_{ij}^\mathrm{E} &= 2\etaS E_{ij} + (\etaD-\etaS) \, \delta_{ij}E_{kk} \, , \\
    \sigma_{ij}^\mathrm{O} &= \etaO
\left(
\epsilon_{ik}E_{kj}
+
\epsilon_{jk}E_{ki}
\right) ,
\end{align}
\end{subequations}
with $\delta_{ij}$ denoting the Kronecker delta and $\epsilon_{ij}$ the 2D Levi-Civita permutation tensor, an antisymmetric tensor defined by $\epsilon_{xx}=\epsilon_{yy}=0$ and $\epsilon_{xy}=-\epsilon_{yx}=1$.
Here, $\etaS$, $\etaD$, and $\etaO$ denote the 2D shear, dilatational, and odd viscosities, respectively.
The shear viscosity $\etaS$ characterizes resistance to shape-changing deformations, the dilatational viscosity $\etaD$ quantifies dissipation associated with isotropic compression or expansion, and the odd viscosity $\etaO$ represents a nondissipative, parity-odd contribution that couples stresses perpendicular to velocity gradients.
In addition, $E_{ij} = (\partial_i v_j + \partial_j v_i)/2$ denotes the symmetric rate-of-strain tensor, while $\sigma_{ij}^\mathrm{E}$ and $\sigma_{ij}^\mathrm{O}$ denote the even and odd contributions to the stress tensor, respectively.

From the momentum balance equation~\eqref{eq:momentum_2D}, together with Eqs.~\eqref{eq:bulk_force} and \eqref{eq:stress_tensor}, the hydrodynamic equation for a 2D compressible fluid is derived as~\cite{hosaka2021nonreciprocal}
\begin{equation}
\boldsymbol{\eta}
\cdot \boldnabla_\parallel^2 \bm{v}
+ \etaD\boldnabla_\parallel (\boldnabla_\parallel\cdot\bm{v})  
-\frac{\etaB}{h} \, \bm{v}
-\frac{h}{2} \, \boldnabla_\parallel p
+\bm{f} = \bm{0} \, ,
\label{eq:governing_eqn_final}
\end{equation}
where we have defined the 2D viscosity tensor
\begin{equation}
    \boldsymbol{\eta} = \etaS \mathds{1}
+ \etaO\boldepsilon \, ,
\end{equation}
where $ \mathds{1}$ denotes the 2D identity tensor.
Our goal in this work is to derive an exact analytical solution to Eq.~\eqref{eq:governing_eqn_final} for a point-force singularity acting on the 2D medium.

Integrating the 3D divergence of the bulk fluid velocity $\bm{u}(\bm{r}, z)$ over the thin lubrication layer for $z \in [0, h]$ and setting it to zero to enforce incompressibility of the underlying fluid gives the relation
\begin{equation}
    \boldnabla_\parallel\cdot\bm{v} = \frac{h^2}{6\etaB}\boldnabla_\parallel^2p \, .
    \label{eq:compressibility}
\end{equation}
This indicates that the 2D fluid is compressible due to the incompressibility constraint of the 3D bulk fluid.
For additional theoretical details on the model and the derivation of the governing equations, we refer the reader to the references~\onlinecite{barentin1999, elfring2016surface}.

For convenience, we introduce two hydrodynamic inverse screening lengths,
\begin{equation}
\kappa = \sqrt{\frac{\etaB}{h \etaS}} \, , \qquad 
\lambda = \sqrt{\frac{2\etaB}{h\bar{\eta}}} \, ,
\label{kappa_lambda}
\end{equation}
which determine the length scale, beyond which the fluid layer transfers momentum to the bulk fluid.
Here, we have defined the mean even viscosity as
\begin{equation}
    \bar{\eta}= 
    \frac{1}{2}
    \left( \etaS+\etaD \right) \, .
\end{equation}

\section{Green's functions in 2D Fourier space}

The hydrodynamic response of the thin fluid layer is described by the Green’s function which connects the force applied to the fluid with the resulting flow velocity.
For a 2D point-force singularity $\bm{f}=\bm{F}\delta(\bm{r}-\bm{r}_0)$ applied at position~$\bm{r}_0$ in the 2D fluid layer, the induced velocity field is expressed as
\begin{equation}
    \bm{v}(\bm{r}) = \frac{1}{\etaS} \,
    \G (\bm{r}) \cdot \bm{F}  \, ,
    \label{eq:Greens_fun_V}
\end{equation}
with $\G(\bm{r})$ denoting the velocity Green’s function, which is a 2D tensor.
The corresponding solution for the pressure field is given by
\begin{equation}
     p(\bm{r}) = \frac{\kappa}{h} \,
    \P (\bm{r}) \cdot \bm{F}  \, , \label{eq:Greens_fun_P} 
\end{equation}
with $\P(\bm{r})$ denoting the pressure Green’s function, which is a 2D vector.

Due to the 2D geometry of the fluid layer, the Green’s function is conveniently obtained using 2D Fourier transforms. This method has been widely used in low Reynolds number hydrodynamics to solve a variety of flow problems in the viscously dominated regime~\cite{felderhof2009flow, bickel2007hindered, bickel2019hydrodynamic, daddi2016long, daddi2016hydrodynamic, daddi2018brownian, daddi2018hydrodynamic, daddi2019frequency}.

We proceed by transforming Eqs.~\eqref{eq:governing_eqn_final} and~\eqref{eq:compressibility} into Fourier space, mapping the spatial position vector $\bm{r}$ to its corresponding 2D wavevector $\bm{k}=(k_x, k_y)$, and denoting the transformed quantities with a tilde.
We denote the magnitude of $\bm{k}$ by $k = |\bm{k}|$, and the phase angle by $\phi = \arctan(k_y/k_x)$.
For convenience, we introduce two orthogonal unit vectors, $ \bm{k}_\parallel = \bm{k}/k $ and $ \bm{k}_\perp = (-k_y, k_x)/k $, oriented parallel and perpendicular to the wavevector, respectively.

In the 2D Fourier space, Eq.~\eqref{eq:Greens_fun_V} takes the form
\begin{equation}
    \widetilde{\bm{v}}(\bm{k}) = \frac{1}{\etaS} \,
    \widetilde{\G}(\bm{k}) \cdot \bm{F} \, .
\end{equation}
The detailed derivation of the Green’s functions in Fourier space can be found in Ref.~\onlinecite{hosaka2021nonreciprocal}. Here, we directly present the expressions and focus on the inverse Fourier transform, which constitutes the main contribution of this paper.
The Fourier-transformed Green's function is obtained as
\begin{equation}
\widetilde{\G} (\bm{k})
= \frac{\etaS\left(k^{2}+\kappa^{2}\right) \mathds{P}_\parallel + 2\bar{\eta}\left(k^{2}+\lambda^{2}\right) \mathds{P}_\perp-\etaO k^{2} \boldepsilon }{2 \bar{\eta}\left(k^{2} + 
\kappa^{2}\right)\left(k^{2}+\lambda^{2}\right)+\etaO^{2} \etaS^{-1} k^{4}} \, , 
\label{eq:Green_function_Fourier}
\end{equation}
with the dyadic projectors $\mathds{P}_\parallel = \bm{k}_\parallel \, \bm{k}_\parallel$ and $\mathds{P}_\perp = \bm{k}_\perp \,  \bm{k}_\perp$, which provide an orthogonal decomposition of the identity in 2D, namely
$\mathds{P}_\parallel + \mathds{P}_\perp = \mathds{1}$, with $\mathds{P}_\parallel \mathds{P}_\perp = \mathbf{0}$.

By rescaling lengths with $\kappa^{-1}$ in Eq.~\eqref{eq:Green_function_Fourier} and defining the scaled wavenumber via $u = k/\kappa$, the Green’s function in Eq.~\eqref{eq:Green_function_Fourier} can be written as
\begin{equation}
    \widetilde{\G} (u,\phi) =
    \frac{5\left( \alpha u^2+1 \right) \mathds{1}
    -3\left( \beta u^2+1\right)
    \mathds{R}_\phi
    -2\mu u^2 \boldepsilon}
    { 2\kappa^2 \left[ 4 \left( u^2+1 \right) \left( \xi^2 u^2+1\right) + \mu^2 u^4 \right] } \, ,
    \label{eq:Green_function_Fourier_scaled}
\end{equation}
where we have defined
\begin{equation}
    \mu = \frac{\etaO}{\etaS } \, , 
\end{equation}
representing the scaled odd viscosity, and
\begin{equation}
    \xi = \frac{\kappa}{\lambda} 
    = \frac{1}{2} \, \sqrt{1+\frac{\etaD}{\etaS} } \, ,
\end{equation}
representing the viscosity ratio, quantifying the contribution of the dilatational viscosity to the resistance to deformation relative to shear.
In addition, the tensor
\begin{equation}
    \mathds{R}_\phi
    = 2 \mathds{P}_\parallel - \mathds{1}
    = \begin{pmatrix}
        \cos (2\phi) & \sin (2\phi) \\[3pt]
        \sin (2\phi) & -\cos (2\phi)
    \end{pmatrix},
    \label{eq:Householder_reflection}
\end{equation}
represents a 2D Householder reflection across the line oriented at an angle $\phi$ with respect to the $x$-axis.
The dimensionless numbers $\alpha$ and $\beta$ appearing in Eq.~\eqref{eq:Green_function_Fourier_scaled} are defined as 
\begin{subequations}
    \begin{align}
    \alpha &= \frac{1}{5} \left( 4\xi^2+1 \right) 
    = \frac{1}{5} \left( 2+\frac{\etaD}{\etaS} \right) , \\
    \beta &= \frac{1}{3} \left( 4\xi^2-1 \right) = \frac {\etaD}{3 \etaS} .
\end{align}
\end{subequations}
Hence, note that $\xi \ge 1/2$, $\alpha \ge 2/5$, and $\beta \ge 0$.

Similarly, the Green’s function for the hydrodynamic pressure in 2D Fourier space can be expressed as
\begin{equation}
    \widetilde{\P}(u,\phi) = 
    -\frac{6i}{ \kappa^2 u}
    \frac{  (u^2+1) \, \bm{k}_\parallel - \mu u^2\, \bm{k}_\perp }{4 \left( u^2+1 \right) \left( \xi^2 u^2+1\right) + \mu^2 u^4} \, ,
    \label{eq:pressure_Fourier_scaled}
\end{equation}
noting that $\bm{k}_\parallel$ and $\bm{k}_\perp$ depend on the polar angle $\phi$.

In the next section, we derive the expressions for the velocity and pressure fields in the real space via the inverse 2D Fourier transform.
For convenience, a summary of the mathematical background on inverse Fourier transform in the system of polar coordinates is provided in the Appendix~\ref{app:polar}.

\section{Green's functions in 2D real space}

\subsection{Velocity field}

Upon taking the inverse 2D Fourier transform of Eq.~\eqref{eq:Green_function_Fourier_scaled}, the velocity Green's function in real-space is obtained as
\begin{equation}
    \G (\R) = 
    Q_0 \mathds{1}
    + Q_1 \mathds{R}_\theta 
    + Q_2 \boldepsilon \, ,
    \label{eq:realG}
\end{equation}
where we have defined 
\begin{subequations} \label{eq:Q1Q2Q3}
    \begin{align}
    Q_0 &= \frac{5}{2} \left( \Phi_0^0 + \alpha \Phi_0^1 \right) , \\
    Q_1 &= \frac{3}{2} \left( \Phi_1^0 + \beta \Phi_1^1 \right) , \\[3pt]
    Q_2 &= - \mu \Phi_0^1 \, .
\end{align}
\end{subequations}
Here, $\Phi_m^q$ with $m,q \in \{0,1\}$ denote the improper integrals defined by
\begin{equation}
    \Phi_{n}^q  = \frac{1}{2\pi} \int_0^\infty
    \mathrm{d}u \, 
    \frac{u^{2q+1} J_{2n} (\rho u) }
    {4 \left( u^2+1 \right) \left( \xi^2 u^2+1\right) + \mu^2 u^4} \, , 
    \label{eq:phi_n_q}
\end{equation}
with $\rho = \kappa r$ denoting the scaled polar distance, and $J_{2n}$ is the Bessel function of the first kind of order $2n$.
Note that $\mathds{R}_\theta$ is a 2D Householder reflection defined by Eq.~\eqref{eq:Householder_reflection}, with $\theta = \arctan(y/x)$ denoting the polar angle in the real space

The integral given by Eq.~\eqref{eq:phi_n_q} is convergent and an analytical evaluation is possible using the method of residues. 
As detailed in Appendix~\ref{app:residues}, the coefficients $Q_n$ for $n \in \{0,1,2\}$ defined in Eqs.~\eqref{eq:Q1Q2Q3} can be written in the form
\begin{equation}
    \label{eq:Qn}
    Q_n = \frac{i}{32 \sqrt{\delta}} \left( \mathcal{Q}_n^- - \mathcal{Q}_n^+ \right)
    + \frac{3}{8\pi \rho^2} \, \delta_{n 1} \, , 
\end{equation}
where
\begin{subequations}
    \label{eq:Q}
    \begin{align}
    \mathcal{Q}_0^\pm &= a_\pm
    H_0^{(1)} \left( i\rho A_\pm \right) , \\
    \mathcal{Q}_1^\pm &= b_\pm
    H_2^{(1)} \left( i\rho A_\pm \right) , \\
    \mathcal{Q}_2^\pm &= 2\mu 
    A_\pm^2 \, H_0^{(1)} \left( i\rho A_\pm \right).
    \label{eq:Q2}
\end{align}
\end{subequations}
In the above, the abbreviations are given by
\begin{equation}
    a_\pm = 5 \left( 1-\alpha A_\pm^2 \right) , \quad
    b_\pm = 3 \left( 1-\beta A_\pm^2 \right) ,
    \label{eq:a_pm_b_pm}
\end{equation}
where
\begin{align}
    A_\pm &= \sqrt{ \frac{2 \left( 1+\xi^2 \pm 
    \sqrt{\delta} \right)}{4 \xi^2 + \mu^2 } } \, , 
    \label{eq:Apm}
    \\
    \delta &= \left( 1-\xi^2\right)^2-\mu^2 \, .
    \label{eq:delta}
\end{align}
Note that $\sqrt{\cdot}$ denotes the principal value of the square root representing the branch with nonnegative real part.

Expression~\eqref{eq:Qn} together with Eqs.~\eqref{eq:Q} represents the central result of this paper.
Equations~\eqref{eq:Q} show that only $\mathcal{Q}_2^\pm$, associated with the antisymmetric part, is an odd function of the odd viscosity, whereas the symmetric parts $\mathcal{Q}_0^\pm$ and $\mathcal{Q}_1^\pm$ depend quadratically on $\mu$; see the definition of $A_\pm$ in Eq.~\eqref{eq:Apm}. This reflects the fundamental symmetry relation commonly observed in odd fluids~\cite{hosaka2021nonreciprocal, khain2022, lier2023lift, yuan2023stokesian, everts2024dissipative}
\begin{align}
    \G(\bm{r},\mu)=\G(\bm{r},-\mu)^\top \, , 
\end{align}
with $\top$ denoting the matrix transpose.
This relation is known as the Onsager--Casimir reciprocity for systems with external sources of time-reversal breaking~\cite{fruchart2023odd}.
Within the specific framework of low-Reynolds-number hydrodynamics, the resulting symmetries in the Green's functions can be explicitly derived via the Lorentz reciprocal theorem~\cite{masoud2019}, which has recently been generalized to fluids odd viscosity~\cite{hosaka2023lorentz}.

In the case $\mu=0$ and $\xi=1$ (equivalently, $\lambda=\kappa$), the solution for the velocity Green's function can be expressed in terms of the modified Bessel function of the second kind. Using the connection formula valid for $\rho>0$,
\begin{equation}
    Y_\nu(i\rho) = i^{\nu+1} I_\nu(\rho) - \frac{2}{\pi} \, i^{-\nu} K_\nu(\rho) \, , 
\end{equation}
with $I_\nu$ and $K_\nu$ denoting modified Bessel functions of the first and second kind, respectively,
we obtain
\begin{equation}
    \G (\rho, \theta) = \frac{3}{16\pi} 
    \left[  \frac{5}{3} \, K_0(\rho) \, \mathds{1} + \left( \frac{2}{\rho^2} - K_2(\rho) \right) \mathds{R}_\theta \right],
    \label{eq:Green_mu0_xi1}
\end{equation}
which is consistent with previously reported results for the Green’s function of a 2D fluid layer in the absence of odd viscosity~\cite{hosaka2021nonreciprocal}.

The velocity Green’s function in Eq.~\eqref{eq:Green_mu0_xi1} can be expressed in the far field $\rho \gg 1$ as
\begin{equation}
    \G(\rho,\theta) \approx
    \frac{3}{8\pi \rho^2} \, \mathds{R}_\theta \, .
    \label{eq:Glarge}
\end{equation}
On the other hand, it can be expressed in the near field $\rho \ll 1$ as
\begin{equation}
    \G(\rho, \theta) \approx \frac{3}{32\pi}
    \left[ 
    \frac{10}{3} \left( \ln \frac{2}{\rho}-\gamma \right) \mathds{1}
    + \mathds{R}_\theta \right],
    \label{eq:Gsmall}
\end{equation}
wherein $\gamma \approx 0.5772$ denotes the Euler--Mascheroni constant.
Asymptotic expressions of the velocity Green’s function were previously obtained in Ref.~\onlinecite{hosaka2023hydrodynamics} in the limit of small odd viscosity relative to the even viscosities, and coincide with Eqs.~\eqref{eq:Glarge} and \eqref{eq:Gsmall} when setting $\mu=0$.

\begin{figure}
    \centering
    \includegraphics[width=0.75\linewidth]{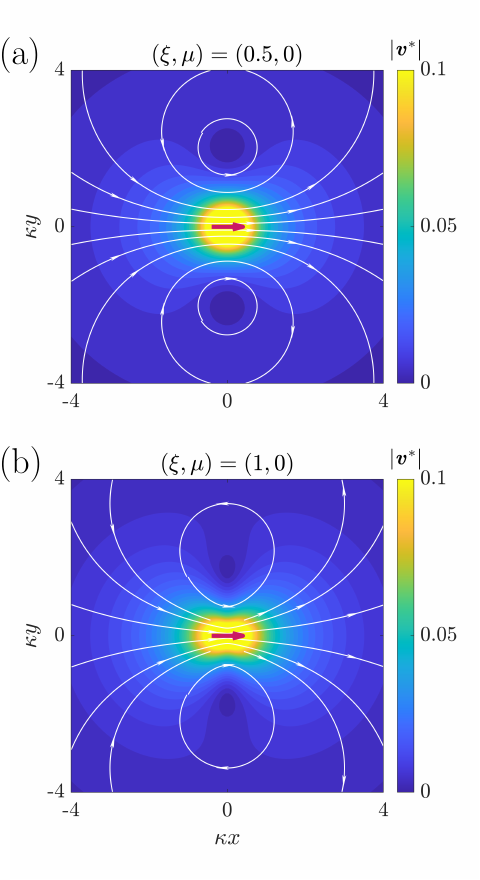}
    \caption{Contour plot of the velocity magnitude with superimposed quiver plot of the 2D flow field induced by a point-force singularity in a compressible fluid layer without odd viscosity ($\mu=0$). Results are shown for (a)~$\xi=0.5$ and (b)~$\xi=1$. The scaled velocity is defined as $\bm{v}^*=(\etaS/F)\bm{v}$, where $F$ denotes the magnitude of the applied point force.}
    \label{fig:monopole1}
\end{figure}

In Fig.~\ref{fig:monopole1} we present representative flow fields generated by a monopole singularity acting at the origin in a compressible fluid layer without odd viscosity ($\mu=0$). Results are shown for two values of the viscosity ratio parameter: (a) $\xi=0.5$, the minimum admissible value of $\xi$, and (b) $\xi=1$. In both cases, the flow is symmetric about the axis aligned with the applied point force and features a pair of symmetrically placed eddies relative to the force location. The eddies counter-rotate in both configurations.
For $\xi=1$ [Fig.~\ref{fig:monopole1}~(b)], the flow is stretched along the $x$-direction, which is the direction of the applied point force.
The pair of vortices is also reminiscent of flows in porous media with a frictional or resistive layer, where momentum screening gives rise to confined, counter-rotating flow structures~\cite{vilfan2025stokes}.
The qualitative changes in the flow structure, even in the absence of odd viscosity, highlight the complex dynamics arising from viscosity contrast.

\begin{figure}
    \centering
    \includegraphics[width=0.75\linewidth]{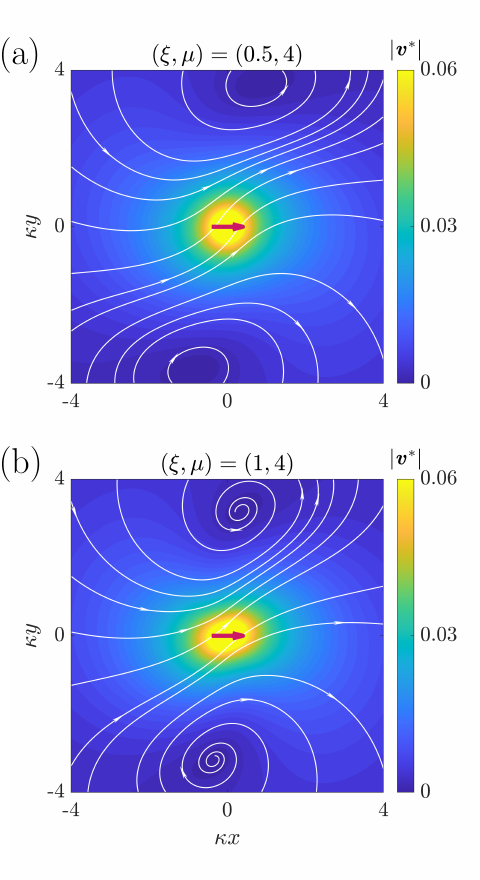}
    \caption{Contour plot of the velocity magnitude with a superimposed quiver representation of the 2D flow field induced by a point-force singularity in a compressible fluid layer with odd viscosity ($\mu=4$). Panel (a)~corresponds to $\xi=0.5$ and panel~(b) to $\xi=1$. The scaled velocity is defined as $\bm{v}^*=(\etaS/F)\bm{v}$.
}
    \label{fig:monopole2}
\end{figure}

Figure~\ref{fig:monopole2} shows contour and quiver plots of the flow induced by a force monopole acting along the $x$-direction in a 2D fluid layer with odd viscosity $\mu=4$, for two values of the viscosity ratio parameter: (a)~$\xi=0.5$ and (b)~$\xi=1$, as before.
The presence of finite odd viscosity breaks the flow symmetry, shifting the positions of the eddy centers.
This symmetry breaking can be understood through the transverse stress response induced by the odd viscosity. Due to the antisymmetric tensor $\bm\epsilon$ [see Eq.~\eqref{eq:governing_eqn_final}], the induced flow is rotated clockwise by $\pi/2$ compared to the conventional flow where such symmetries are conserved $(\mu=0)$. A qualitative change in the flow structure is observed when varying $\xi$.
For $\xi = 0.5$, the flow organizes into closed, circular eddies that resemble the swirling patterns observed when odd viscosity is absent [Fig.~\ref{fig:monopole2}~(a)]. In contrast, for $\xi = 1$, the eddies no longer remain closed but instead show converging or diverging behavior, indicating a significant alteration of the flow structure due to the increased influence of odd viscosity [Fig.~\ref{fig:monopole2}~(b)]. This difference originates from the interplay between odd-viscous stresses and the viscosity ratio, which modifies the balance between rotational (solenoidal) and compressional (irrotational) flow components. In the presence of odd viscosity and a viscosity ratio, these two components become coupled. Odd viscosity converts part of the compressional response induced by the point force into rotational motion, while the viscosity ratio controls the relative magnitude and spatial structure of the two contributions.
This balance determines whether the flow organizes into closed vortical eddies or into converging and diverging patterns.

\subsection{Pressure field}

The pressure can be derived directly in the real space by applying the inverse Fourier transform to Eq.~\eqref{eq:pressure_Fourier_scaled}, involving the evaluation of another set of improper integrals that can also be handled using the method of residues. 

The pressure vector can be expressed in the real space, in polar coordinates, as
\begin{equation}
    \P (\R) = 
    P_\rho \, \hat{\bm{e}}_\rho + 
    P_\theta \, \hat{\bm{e}}_\theta \, ,
\end{equation}
where $\hat{\bm{e}}_\rho$ and $\hat{\bm{e}}_\theta$ are the unit basis vectors of the polar coordinate system.
In addition,
\begin{equation}
    P_\rho = \Psi^0+\Psi^1  \, , \quad
    P_\theta = - \mu \Psi^1 \, .
\end{equation}
Here, we have introduced the series of improper integrals
\begin{equation}
    \Psi^q =
    \frac{3}{\pi}
    \int_0^\infty
    \mathrm{d}u \, 
    \frac{u^{2q} J_1 (\rho u) }
    {4 \left( u^2+1 \right) \left( \xi^2 u^2+1\right) + \mu^2 u^4} \, , 
    \label{eq:psi_q}
\end{equation}
where $q \in \{0,1\}$.

Likewise, the integral defining $\Psi^q$ is convergent, allowing for an analytical evaluation by using the method of residues.
As for the integrals associated with the velocity field, we define a function in the complex plane and evaluate it by integration along a contour, as shown in Fig.~\ref{fig:contour_integration}.
We obtain
\begin{equation}
    \Psi^q = H^q(A_-) - H^q(A_+) + \frac{3}{4\pi\rho}\, \delta_{q,0} \, , 
\end{equation}
where $A_\pm$ are defined in Eq.~\eqref{eq:Apm} and
\begin{equation}
    H^q(\zeta) = \frac{3}{8\sqrt{\delta}} \, (-1)^q \, \zeta^{2q-1} H_1^{(1)}(i\rho \zeta) \, ,
\end{equation}
where $\delta$ was defined in Eq.~\eqref{eq:delta}.

Combining these results and using the integral expression given above, $P_i$ for $i \in \{\rho, \theta\}$ can be cast in the form
\begin{equation}
    P_i = \mathcal{P}_i^- - \mathcal{P}_i^+ + \frac{3}{4\pi\rho} \, \delta_{i\rho} \, , 
\end{equation}
where 
\begin{subequations}
    \begin{align}
    \mathcal{P}_\rho^\pm &=
     \frac{3}{8\sqrt{\delta}}
    \left( A_\pm^{-1} - A_\pm \right) 
    H_1^{(1)} \left( i\rho A_\pm \right) , \\[3pt]
    \mathcal{P}_\theta^\pm &= 
     \frac{3 \mu}{8\sqrt{\delta}} \, 
     A_\pm \, H_1^{(1)} \left( i\rho A_\pm \right) \, .
\end{align}
\end{subequations}

For $\mu=0$ and $\xi=1$ (equivalently, $\lambda=\kappa$), the pressure vector reduces to
\begin{equation}
    \P (\rho, \theta) = \frac{3}{4\pi}
    \left( \frac{1}{\rho} - K_1(\rho) \right) \hat{\bm{e}}_\rho \, .
    \label{eq:P_mu0_xi1}
\end{equation}

In the far field $\rho \gg 1$, Eq.~\eqref{eq:P_mu0_xi1} can be expressed as
\begin{equation}
    \P (\rho, \theta) \approx \frac{3}{4\pi \rho} \, .
\end{equation}
In the near field $\rho \ll 1$, accurate up to $\mathcal{O}\left( \rho^3 \right)$, it is given by
\begin{equation}
    \P (\rho, \theta) \approx \frac{3 \rho}{8\pi}
    \left( \ln \frac{2}{\rho}-\gamma+\frac{1}{2} \right) \hat{\bm{e}}_\rho \, .
\end{equation}

\begin{figure}
    \centering
    \includegraphics[width=0.75\linewidth]{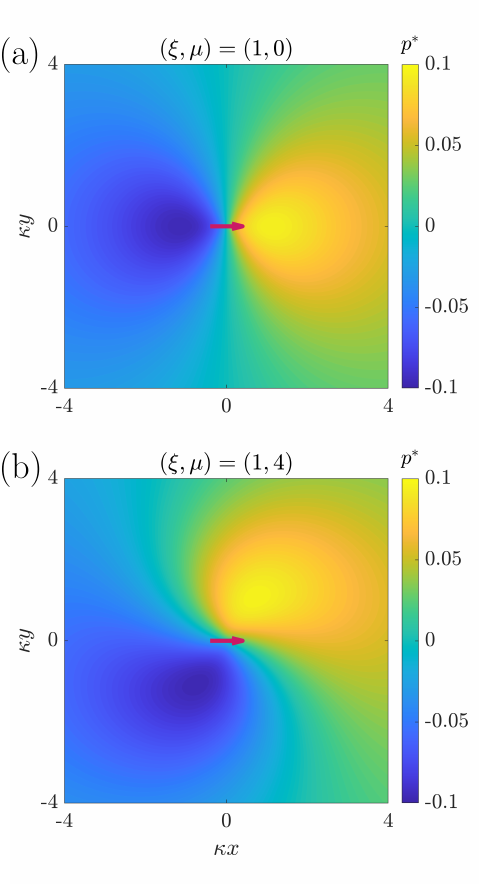}
    \caption{Contour plot of the pressure field induced by a point force in a 2D fluid: (a)~without odd viscosity ($\mu=0$) and (b)~with odd viscosity ($\mu=4$). Results are shown for $\xi=1$. The scaled pressure is defined as $p* = h/\left(\kappa F\right) p$.
    }
    \label{fig:pressure}
\end{figure}

In Fig.~\ref{fig:pressure} we show the pressure field in an odd fluid layer with viscosity ratio parameter $\xi=1$ for (a)~$\mu=0$ and (b)~$\mu=4$. In the absence of odd viscosity, the pressure field is symmetric about the $x$-axis, with negative pressure for $x<0$ and positive pressure for $x>0$ [Fig.~\ref{fig:pressure}~(a)]. Including odd viscosity strongly distorts the pressure field and breaks this symmetry [Fig.~\ref{fig:pressure}~(b)].
This asymmetric pressure distribution is directly relevant for force transmission to embedded or nearby objects and can lead to qualitatively different hydrodynamic interactions and transport behaviors.

\section{Dipolar flow field}

The exact analytical calculations presented in the previous section provide the foundation for determining higher-order singularities. Here, we focus on the leading-order contribution, namely the force dipole, which is obtained by placing two force monopoles infinitesimally close to each other. This flow represents the velocity field generated by various self-propelling active microswimmers, including bacteria and algae~\cite{lauga2016bacterial}.

We define the unit vector $\hat{\bm{e}}_\parallel =(\cos\varphi,\sin\varphi)^\top$ to represent the orientation of the force dipole with respect to the $x$-axis and denote by $\sigma$ the strength of the dipole singularity. The dipole flow field can be written as
\begin{equation}
    \bm{v}_\mathrm{D} = -\sigma 
    \left( \hat{\bm{e}}_\parallel \cdot \boldnabla_\parallel \right)
    \left( \G \cdot \hat{\bm{e}}_\parallel \right) .
\end{equation}

Defining the unit vectors $\hat{\bm{e}}_\perp = (-\sin\varphi, \cos\varphi)^\top$ and $\hat{\bm{u}}_n = (\cos\vartheta_n, \sin\vartheta_n)^\top$, with $\vartheta_n = (n+1)\theta - n\varphi$ for $n \in \{1,2\}$, the dipole flow field can be written in the compact form as
\begin{equation}
    \bm{v}_\mathrm{D}(\R) = \kappa\sigma
    \left[ 
    \bm{\mathcal{U}}
    +\bm{\mathcal{S}}_- - \bm{\mathcal{S}}_+
    +\left( \bm{\mathcal{C}}_- - \bm{\mathcal{C}}_+ \right) \cos(\theta-\varphi)
    \right] , 
\end{equation}
where
\begin{equation}
    \bm{\mathcal{U}} = \frac{3}{4\pi\rho^3}\,\hat{\bm{u}}_2 \, .
\end{equation}
In addition, 
\begin{equation}
    \bm{\mathcal{S}}_\pm = \frac{2i}{32\sqrt{\delta} \rho} \,
    b_\pm H_2^{(1)}(i\rho A_\pm) \, \hat{\bm{u}}_2 \, , 
\end{equation}
and 
\begin{equation}
    \bm{\mathcal{C}}_\pm = \frac{A_\pm}{32\sqrt{\delta}}
    \left( 
    b_\pm \hat{\bm{u}}_1
    -a_\pm \hat{\bm{e}}_\parallel
    +2\mu A_\pm^2 \hat{\bm{e}}_\perp
    \right)  H_1^{(1)}(i\rho A_\pm) \, , 
\end{equation}
where $a_\pm$ and $b_\pm$ were defined earlier in Eq.~\eqref{eq:a_pm_b_pm}.

In the isotropic case with $\xi=1$ and $\mu=0$, the dipole flow field simplifies to
\begin{equation}
    \bm{v}_\mathrm{D} (\rho,\theta)  = \frac{3\kappa\sigma}{16\pi}
    \left[ 
    K_1(\rho) \, \hat{\bm{w}}
    +\frac{2}{\rho} \left( \frac{2}{\rho^2} - K_2(\rho) \right) \hat{\bm{u}}_2
    \right] ,
    \label{eq:dipole_mu0_xi1}
\end{equation}
where we have defined
\begin{equation}
    \hat{\bm{w}} = 
    \left( \frac{5}{3}\, \hat{\bm{e}}_\parallel - \hat{\bm{u}}_1 \right) \cos(\theta-\varphi) \, .
\end{equation}

In the far field $\rho \gg 1$, Eq.~\eqref{eq:dipole_mu0_xi1} can be expressed as
\begin{equation}
    \bm{v}_\mathrm{D}(\rho,\theta) \approx
    \frac{3\kappa\sigma}{4\pi\rho^3} \, .
    \label{eq:vDlarge}
\end{equation}
In the near-field limit $\rho \ll 1$, we obtain
\begin{equation}
    \bm{v}_\mathrm{D} (\rho,\theta) \approx  \frac{3\kappa\sigma}{16\pi \rho}
    \left( \hat{\bm{w}}  + \hat{\bm{u}}_2 
    \right).
    \label{eq:vDsmall}
\end{equation}

\begin{figure}
    \centering
    \includegraphics[width=0.75\linewidth]{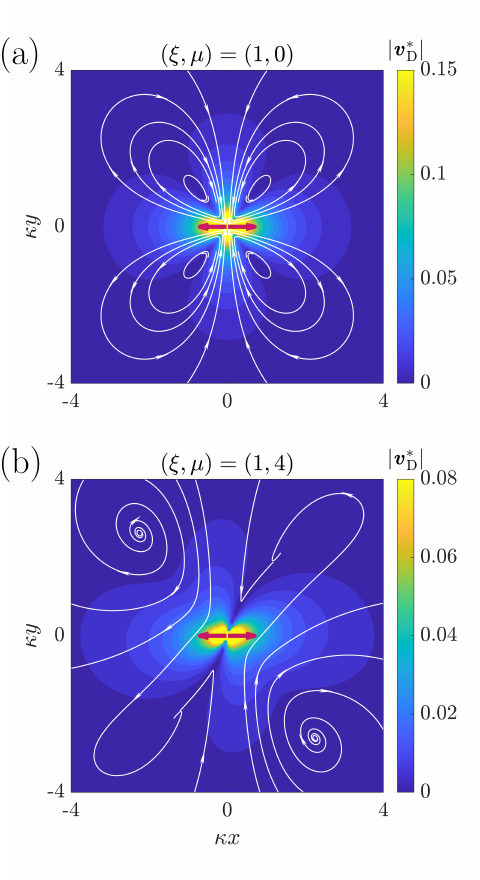}
    \caption{Contour and quiver plots of the fluid velocity induced by a force dipole aligned along the $x$ axis: (a)~without odd viscosity and (b)~with odd viscosity $(\mu=4)$. Results are shown for $\xi=1$.
    The scaled dipolar velocity is defined as $\bm{v}_\mathrm{D}^* = \bm{v}_\mathrm{D}/\left(\kappa\sigma\right)$.}
    \label{fig:dipole}
\end{figure}

In Fig.~\ref{fig:dipole}, we show the contour plots together with a quiver representation of the flow field induced by a force dipole in a 2D odd fluid layer. The results are presented for $\sigma>0$, corresponding to pusher (extensile) swimmers. The flow generated by a puller (contractile) swimmer is identical, except that the direction of the flow is reversed. We fix the viscosity ratio parameter to $\xi=1$ and vary the odd viscosity parameter $\mu$, with (a)~$\mu=0$ and (b)~$\mu=4$. In the absence of odd viscosity, the flow is symmetric with respect to the direction along which the force dipole is applied and exhibits butterfly-shaped vortical structures [Fig.~\ref{fig:dipole}~(a)]. For nonzero odd viscosity, the flow is strongly distorted, characterized by the emergence of converging vortices and a loss of symmetry with respect to the force-dipole direction [Fig.~\ref{fig:dipole}~(b)]. Overall, these modifications of the dipolar flow field are expected to have profound consequences for the hydrodynamic interactions, orientational dynamics, and collective behavior of swimming microorganisms confined to fluid layers with odd viscosity.

\section{Summary and Discussion}

In summary, we have derived exact analytical expressions for the linear response functions of a 2D compressible fluid layer with odd viscosity, together with conventional shear and dilational viscosities. These results provide a complete description of the hydrodynamic fields generated by monopole and dipole singularities, highlighting the emergence of transverse flows and vortical patterns driven by the antisymmetric stress contributions from odd viscosity. The interplay between the 2D fluid layer and the underlying 3D hydrodynamics gives rise to rich flow structures, with hydrodynamic screening lengths determined solely by the even viscosities and the transverse responses governed exclusively by the odd-viscosity coefficient.

Momentum in a fluid layer is conserved over distances smaller than the hydrodynamic screening length, but beyond this length scale, it leaks into the surrounding fluid~\cite{diamant2009hydrodynamic}.
As shown in Eq.~\eqref{eq:Glarge}, the velocity Green's function decays as $1/\rho^2$ at large distances $\rho \gg 1$. This algebraic decay arises because mass is conserved in two dimensions, whereas total momentum is not conserved due to the presence of an underlying rigid substrate.
At small scales $\rho \ll 1$, the velocity instead decays logarithmically, as in Eq.~\eqref{eq:Gsmall}, consistent with 2D momentum conservation. The behaviors of higher-order singularities can be inferred similarly. For instance, a first-order flow singularity is expected to scale as $1/\rho^3$ in the far field ($\rho \gg 1$) and as $1/\rho$ in the near field ($\rho \ll 1$). These limiting behaviors correspond to those of a dipolar flow and agree with Eqs.~\eqref{eq:vDlarge} and \eqref{eq:vDsmall}.
Similar crossover behaviors have also been reported for incompressible fluid layers supported by a substrate~\cite{oppenheimer2010, hosaka2017}.

Our work builds on the broader framework of hydrodynamics in chiral active matter, where the violation of time-reversal and parity symmetries leads to nonreciprocal flow behavior and the emergence of odd viscosity as a novel transport coefficient. The exact Green’s functions derived here serve as a fundamental building block for resistance tensors, multipole expansions, and boundary integral methods~\cite{
masoud2019, pozrikidis1992}, enabling precise predictions of hydrodynamic interactions in compressible odd-viscous fluids. These findings are particularly relevant for understanding the dynamics of microswimmers, driven rotor collectives, and other active matter systems in confined or supported geometries, where the presence of odd viscosity can strongly affect propulsion, alignment, and collective behavior.

Looking ahead, our results provide a foundation for exploring more complex settings, including multi-particle interactions and various types of boundary conditions at the substrate. They also motivate experimental studies aimed at probing the characteristic transverse flows and nonreciprocal responses in engineered odd-viscous fluid layers. By bridging exact theoretical predictions with potential applications in active and chiral fluid systems, this work paves the way for a deeper understanding of transport phenomena in non-equilibrium, symmetry-breaking environments.

\appendix

\section{2D inverse Fourier transformation in polar coordinates}
\label{app:polar}

In this Appendix, we provide the formulas used to compute 2D inverse Fourier transforms in polar coordinates. Detailed treatments and proofs can be found in standard textbooks on Fourier analysis~\cite{bracewell99, folland2009fourier, stein2011fourier}. A compact overview of the relevant techniques is also provided in Ref.~\onlinecite{baddour2011two}.

The 2D inverse Fourier transform of a function expressed in polar coordinates, $\widetilde{f}(k,\phi)$, can be written in terms of a Fourier series as
\begin{equation}
    f(\rho,\theta) 
    = \sum_{n=-\infty}^\infty f_n(\rho) \, e^{in\theta} \, , 
    \label{eq:infFourier_series}
\end{equation}
where
\begin{equation}
\hspace{-0.2cm}
f_n(\rho) = \frac{i^n}{(2\pi)^2} 
\int_0^\infty \mathrm{d} k \, k
\, J_n(\rho k)    
\int_0^{2\pi}  \mathrm{d}\phi \, 
\widetilde{f}(k,\phi) \, e^{-in\phi} \, .
\end{equation}
Here, $J_n$ again denotes the Bessel function of the first kind of order $n$~\cite{abramowitz2000handbook}.

Since the Fourier-transformed velocity field contains terms independent of $\phi$ and proportional to $\mathds{R}_\phi$ [see Eq.~\eqref{eq:Green_function_Fourier_scaled}], only the modes $n=0$ and $n=\pm 2$ survive in the series representation \eqref{eq:infFourier_series}.
For the pressure field [see Eq.~\eqref{eq:pressure_Fourier_scaled}], only the modes $n=\pm 1$ remain.

\section{Derivation of Eq.~\eqref{eq:realG}}
\label{app:residues}

Here we derive the real-space Green's function in Eq.~\eqref{eq:realG} using the method of residues. 
The core idea is to define a function $f(\zeta)$ of a complex variable~$\zeta$ from the integral equation~\eqref{eq:phi_n_q} as
\begin{equation}
    f(\zeta) = 
    \frac{1}{2\pi}
    \cfrac{\zeta^{2q+1} H_{2n}^{(1)}(\rho \zeta)}
    {4 \left( \zeta^2+1 \right) \left( \xi^2 \zeta^2+1\right) + \mu^2 \zeta^4} \, , 
\end{equation}
where $H_{2n}^{(1)}$ is the Hankel function of the first kind of order $2n$, defined as~\cite{abramowitz2000handbook}
\begin{equation}
    H_{2n}^{(1)}(\zeta) = J_{2n}(\zeta) + i Y_{2n}(\zeta) \, ,
\end{equation}
and $Y_{2n}$ is the Bessel function of the second kind of order $2n$.
The function $f(\zeta)$ is meromorphic, being analytic everywhere except at a finite set of poles.

\begin{figure}
    \centering
    \includegraphics[width=\linewidth]{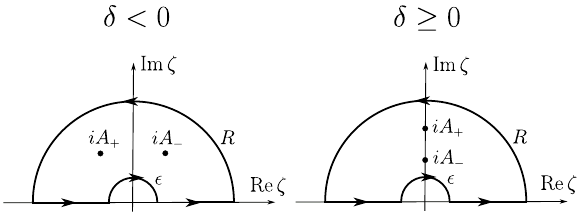}
    \caption{
    Integration contour and the corresponding locations of the residues for $\delta<0$ and $\delta\ge 0$. The contour consists of four segments: two semicircles and two straight lines along the real axis. 
    There are two poles in the upper half-plane, whose positions depend on the sign of $\delta$ defined in Eq.~\eqref{eq:delta}.
    The integrals are evaluated in the limits $\epsilon\to0$ and $R\to\infty$ using the method of residues.}
    \label{fig:contour_integration}
\end{figure}

We integrate $f(\zeta)$ along a contour composed of the following segments: the upper side of the branch cut for $\zeta \in [-R,-\epsilon]$, a small clockwise semicircle of radius $\epsilon$ around the origin of the complex plane, the positive real axis for $\zeta\in [\epsilon, R]$, and finally the upper half of the circle $|\zeta| = R$ traversed counterclockwise; see Fig.~\ref{fig:contour_integration}.
The function is chosen so that the integral over the large semicircle vanishes in the limit $R \to \infty$.
For further technical details on the mathematical steps, we refer the reader to the Appendix of Ref.~\onlinecite{daddi2025hydrodynamic}, where a closely related evaluation is carried out.

In the upper half of the complex plane, the function $f(\zeta)$ has two poles located at $\zeta = iA_\pm$, where 
\begin{equation}
    A_\pm = \sqrt{ \frac{2 \left( 1+\xi^2 \pm 
    \sqrt{\delta} \right)}{4 \xi^2 + \mu^2 } } \, , 
\end{equation}
where $\delta = \left( 1-\xi^2\right)^2-\mu^2$ and $\sqrt{\cdot}$ denotes the principal value of the square root representing the branch with nonnegative real part.

The poles lie in the upper half of the complex plane, as the imaginary part of $\zeta = i A_\pm$ is always positive.
Note that for $\delta < 0$, $A_+$ and $A_-$ form a pair of complex conjugates, whereas for $\delta \ge 0$, $A_\pm$ are real numbers satisfying $A_+\ge A_-$, such that $\zeta=iA_\pm$ are purely imaginary, as shown in Fig.~\ref{fig:contour_integration}.
The integral in Eq.~\eqref{eq:phi_n_q} is thus given by the sum of the residues as
\begin{equation}
    \Phi_{n}^q =  \Gamma_{n}^q(A_-) - \Gamma_{n}^q(A_+) + \frac{1}{4\pi\rho^2} \, \delta_{n1} \delta_{q0} \, , \label{eq:phi_n_q_res}
\end{equation}
where
\begin{equation}
     \Gamma_{n}^q(\zeta) = \frac{i}{16 \sqrt{\delta} } \, (-1)^q \, \zeta^{2q} H_{2n}^{(1)}(i{\rho} \zeta) \, .
\end{equation}
The term containing the Kronecker deltas in Eq.~\eqref{eq:phi_n_q_res} arises from the integration over half of the small circle around the origin, which exists only when $n=1$ and $q=0$.
By combining these results and using the integral expression derived above, we obtain the coefficients, $Q_0,Q_1,$ and $Q_2$ in Eq.~\eqref{eq:realG}.

\section*{Author contribution}
A.D.M.I., Y.H., and S.K. conceived and designed the research, contributed to writing, and reviewed and edited the manuscript.
A.D.M.I. carried out the analytical calculations and prepared the figures.

\begin{acknowledgments}
Y.H.\ acknowledges support from JSPS Overseas Research Fellowships (Grant No.\ 202460086).
S.K.\ acknowledges the support by National Natural Science Foundation of China (No.\ 12274098) and the startup grant of Wenzhou Institute, University of Chinese Academy of Sciences (No.\ WIUCASQD2021041). 
This work was supported by the JSPS Core-to-Core Program ``Advanced core-to-core network for the physics of self-organizing active matter'' (JPJSCCA20230002).
\end{acknowledgments}

\section*{Declaration of interests}
The authors report no conflict of interest.

\vspace{1cm}
\section*{Data availability}
The data that support the findings of this article are not publicly available. The data are available from the corresponding author upon reasonable request.

\section*{Author ORCIDs}
\begin{itemize}
    \item Abdallah Daddi-Moussa-Ider: \href{https://orcid.org/0000-0002-1281-9836}{0000-0002-1281-9836}
    \item Yuto Hosaka: \href{https://orcid.org/0000-0002-6202-4206}{0000-0002-6202-4206}
    \item Shigeyuki Komura: \href{https://orcid.org/0000-0003-3422-5745}{0000-0003-3422-5745}
\end{itemize}

\section*{References}

%


\end{document}